# Error Concealment in Image Communication Using Edge Map Watermarking and Spatial Smoothing

Shabnam Sodagari
Electrical Engineering Dept.
University Park, PA 16803
shabnam@psu.edu

Peyman Hesami, Alireza Nasiri Avanaki
Control and Intelligent Processing Center of Excellence
School of ECE, University of Tehran
p.hesami@ieee.org, avanaki@ut.ac.ir

*Abstract:* We propose a novel error concealment algorithm to be used at the receiver side of a lossy image transmission system. Our algorithm involves hiding the edge map of the original image at the transmitter within itself using a robust watermarking scheme. At the receiver, wherever a lost block is detected, the extracted edge information is used as border constraint for the spatial smoothing employing the intact neighboring blocks in order to conceal errors. Simulation results show the superiority of our technique over existing methods even in case of high packet loss ratios in the communication network.

## I. INTRODUCTION

Multimedia data (e.g., images) transmitted over communication channels and networks are prone to various distortions and noise types that cause them to be received with deviations from the original one sent at the transmitter side.

In case of the occurrence of a transmission error, retransmission is not always a suitable solution because it can lead to heavier traffic within the network, especially when the size of the media is relatively large e.g. image or video.

Therefore, it is inevitable for the receiver to consider some provisions for alleviating the effects of channel imperfections. In this regard, image error concealment techniques aim at improving the image quality at the receiver in spite of the existence of channel noise.

Current image error concealment methods are either based on using the spatial redundancy (e.g. [1-4]) or data hiding [5-8] or both [9].

In this paper, we present a novel image error concealment method. Our scheme is based on embedding the information existing in the edge map of the image within itself using a robust watermarking algorithm, which can remain unchanged even after JPEG compression. Then at the receiver, the watermarked data is extracted and in order to replace lost image blocks, an image smoothing algorithm is applied using the neighboring uncorrupted blocks. The extracted edge information corresponding to the lost block serves as a border constraint for stopping the smoothing process and starting it from a different direction which entails using a different neighboring block to further enhance the reconstruction quality.

The organization of the paper is as follows: in section II the details of our proposed technique are discussed. Implementation results are demonstrated in section III. Finally, conclusions are derived in section IV.

## II. THE PROPOSED ERROR CONCEALMENT METHOD

In our watermark-based image error concealment scheme, we first compute the binary edge map of the original picture at the transmitter. We then divide the image into $8 \times 8$ non-overlapping blocks. The edge information is then hidden within the original image as a robust (to JPEG compression) and invisible watermark. The edge information of each block is embedded in a way that no block carries its own edge map, so that if a block is lost at the receiver, its corresponding edge data is preserved.

The watermarking technique [11] we use for embedding and extracting the binary edge information within the original image is a robust to JPEG compression Quantization Index Modulation (QIM) based scheme [10], which can be extracted at the receiver with no error.

### A. Embedding the Watermark at the Transmitter

1. If $m$ bits are used to represent each pixel, subtract $2^{m-1}$ from the image [12].
2. Divide the image into non-overlapping $8 \times 8$ blocks and compute the Discrete Cosine Transform (DCT) coefficients of each block [12].
3. Quantize the DCT coefficients using common JPEG quantization matrix [12].
4. The computed coefficients in step 3 should be truncated to an integer. In the original JPEG standard this is done by rounding to the nearest integer [12]. We, instead, round each coefficient into its nearest even integer, if the watermark bit is zero (Bit$^i$ = 0), otherwise (i.e., the current watermark bit is one) the coefficient is rounded to its nearest odd integer.
5. Zigzag scan each block and encode it by the commonly used JPEG Huffman tables, but do not encode the DC components using differential coding.

*B. Extracting the Watermark at the Receiver*

The above algorithm is performed in the reverse order in order to extract the compressed binary edge map at the receiver.

Since the watermark visibility of the above mentioned watermarking algorithm is best when the 8 pixels corresponding to largest DCT JPEG quantization coefficients (i.e., *(1,2), (1,3), (2,1), (2,2), (2,3), (3,1), (3,2), (4,1)*) are used for insertion, we have to pick only 8 edge pixels for insertion. Since our spatial smoothing scheme at the receiver is based on the arrangements shown in Figure 1, the best choice is to select the 8 vertical edge pixels almost in the middle of each block (as shown in Figure 2). The original image and the hidden edge map within it are then sent through the channel. At the receiver, we extract the binary edge information. Wherever a lost block is detected its corresponding edge is taken out from the watermarked image. Our proposed error concealment technique at the receiver can be described as follows:

1- For a lost pixel of a lost block we start from the right neighboring pixels and put the average of 3 adjacent pixels in place of lost pixel. Wherever the smoothing algorithm encounters an edge, it is stopped.
2- We repeat this procedure for the lost block this time using the left neighboring blocks.
3- We add the data of step 1 and 2 and insert it in place of the lost block.

### III. IMPLEMENTATION RESUTLS

We implemented our algorithm for several images for a range of network packet loss probabilities spanning 1% to 70%. Figure 3 shows the results of applying our error concealment method for *baboon* with a packet loss probability of 10%. Figure 4 depicts the PSNR's resulted from error concealment for *lena, peppers, zelda and cameraman* for a variety of packet loss probabilities.

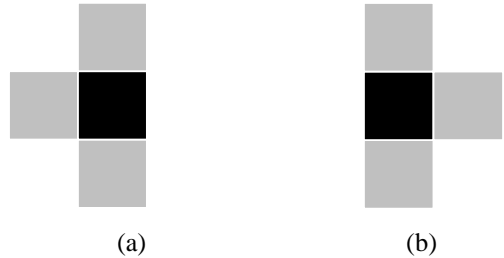

Figure 1. The arrangement of pixels in the spatial smoothing at the receiver

Figure 2. The selected 8 vertical pixels in the edge map of each block which are embedded as a watermark

We have also compared our technique with other common methods and have reported the results in Table 1. Simulations show that our method outperforms the best of the results, which were reported in [9] by almost 1 dB.

### IV. CONCLUSIONS AND FUTURE WORK

An efficient scheme for error concealment at the receiver entity of an image communication system, based on edge map hiding and spatial smoothing, was put forward and its superiority over other existing methods was shown.

The method can be further improved by a more spacious watermarking technique which is also robust to channel noise and has enough capacity for embedding a more detailed edge map. In addition, it can be applied to color images by applying it to each color channel. Using it for video error concealment can be elaborated as a future work.

## V. ACKNOWLEDGEMENT
We would like to thank the authors of [8] who kindly provided us with their simulation codes.

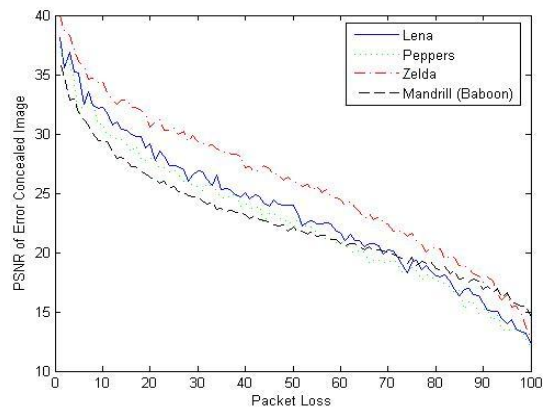

Figure 3. Results of our error concealment method in terms of PSNR vs. packet loss probability for *lena, peppers, zelda and mandrill*

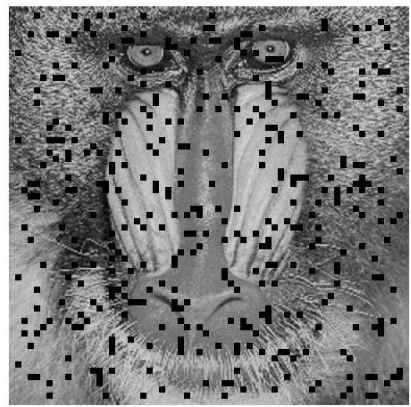

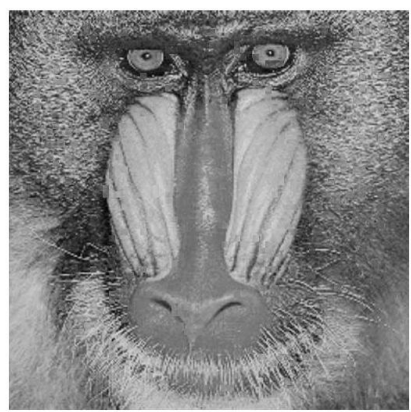

Figure 4. (top) *baboon* with packet loss 10% (bottom) result of applying the proposed error concealment method

Table 1. COMPARISON OF THE EXISTING ERROR CONCEALMENT SCHEMES WITH THE PROPOSED METHOD. THE DATA HAS BEEN OBTAINED FROM [7] AND THE DETAILS OF THE SIMULATION CONDITIONS ARE GIVEN IN [5]

| **Image** | [3] | [1] | [4] | [2] | [5] | [7] | [9] | Our method |
|---|---|---|---|---|---|---|---|---|
| *Lena* | 23.93 | 23.99 | 24.41 | 24.96 | 26.46 | 28.23 | 30.80 | **31.75** |
| *Peppers* | 22.19 | 23.69 | 24.06 | 24.48 | 27.25 | 29.47 | 30.22 | **31.01** |
| *Zelda* | 26.35 | 27.13 | 26.40 | 27.36 | 28.33 | 29.08 | 33.92 | **33.91** |
| *Baboon* | 17.46 | 18.98 | 19.02 | 17.42 | - | 21.92 | 26.92 | **29.13** |
| *Average* | 22.48 | 23.44 | 23.47 | 23.55 | - | 27.17 | 30.46 | **31.45** |